\documentclass[english,prl,twocolumn,groupedaddress,reprint]{revtex4}
\usepackage{graphicx}
\usepackage{amsfonts}
\usepackage[T1]{fontenc}
\usepackage[latin9]{inputenc}
\usepackage{color}
\usepackage{amsmath}
\usepackage{amssymb}
\usepackage{esint}
\usepackage{comment}
\usepackage{bbold}
\usepackage{tikz}
\usepackage{minitoc}
\usepackage{bm}
\usepackage{CJK}

\begin{document}

\title{A classical postselected weak amplification scheme via thermal light cross-Kerr effect}
\author{Tao Wang\footnote{suiyueqiaoqiao@163.com} $^{1,2}$ , Lin-Hai Jiang  $^{3}$, Gang Li  $^{4}$, Chang-Bao Fu $^{1,2}$  and Xue-Mei Su\footnote{ suxm@jlu.edu.cn} $^{1}$}
\affiliation{$^{1}$College of Physics, Jilin University, Changchun 130012, People's Republic of China}
\affiliation{$^{2}$College of Physics, Tonghua Normal University, Tonghua 134000, People's Republic of China}
\affiliation{$^{3}$Analytical and Testing Center, Southwest University of Science and Technology, Mianyang 621010, People's Republic of China}
\affiliation{$^{4}$School of Physics and Optoelectronic Technology,
Dalian University of Technology, Dalian 116024, People's Republic of China}
\date{\today }

\begin{abstract}
In common sense, postselected weak amplification must be related to destructive interference effect of the meter system, and a single photon exerts no effect on thermal field via cross-phase-modulation (XPM) interaction. In this Letter we present, for the first time, a thermal light cross-Kerr effect. Through analysis, we reveal two unexpected results: \emph{i)} postselection and weak amplification can be explained at a classical level without destructive interference, and \emph{ii)} weak amplification and weak value are not one thing.  After postselection a new mixed light can be generated which is nonclassical. This scheme can be realized via electromagnetically-induced transparency.~~~~\newline
~~~~\newline
PACS numbers: 42.65.Dv, 03.65.Ta, 42.50.Hw
\end{abstract}

\maketitle
\emph{Introduction.} - Cross-phase-modulation (XPM) interaction is often used to induce a phase shift (PS) on a beam of coherent light \cite{Knight}. However when a sinlg photon interacts with the coherent light, the PS is difficult to detect. In these years experimental observation of optical nonliearity at the single-photon level have been performed in \cite{Edamatsu09,Gaeta13}. Until recently the nonlinear phase shift due to a postselected single photon was directly measured using electromagnetically-induced transparency and slow light \cite{Steinberg15}. Considering the probe field is in fact not single mode,
Bing He \emph{et al} discussed XPM between a single photon and a coherent light via the continuous mode
treatment of photonic pulse interactions \cite{BingHe11}. There are many interesting phenomena and applications based on XPM, such as quantum nondemolition measurement \cite{Yamamoto85},  protecting the single-photon entangled state \cite{Sheng15} and generating strong micro-macro entanglement \cite{Simon15}.

A thermal state is the most classical state, which is a statistical mixture of coherent
states. Its density matrix $\rho_{th}$ is diagonal in the number
state basis $|n\rangle$, so it can be completely described with number distribution $P(n)$, and does not exhibit  phase  dependence.
In the single-mode approximation, a single photon takes no effect on a thermal light via XPM. Recently in postselected weak measurement (PWM), it is demonstrated theoretically that, thermal state pointer can greatly enhance the postselection success probability, measurement sensitivity and precision  \cite{Li1,Li2}. This inspires us to explore the effect of  single-photon cross-Kerr nonlinearity on thermal light in the PWM scenario.

\begin{figure}[b]
\includegraphics[scale=0.25]{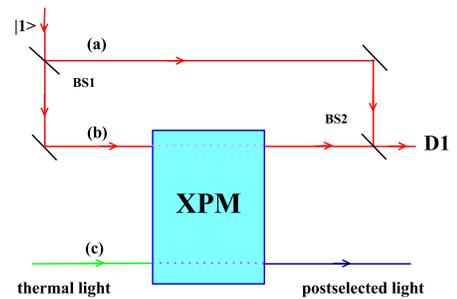}.
\caption{A schematic in our scheme. The single-photon entangled state weakly interacts with the thermal light via XPM. When a photon is detected in the dark port D1, a new mixed state for the probe light can be generated. Here (a),(b) denote the two arms passed by the single-photon, (c) denote the path passed by the probe field. When the single-photon passes the arm (b), it interacts with the thermal light.}
\end{figure}

The idea of PWM proposed by Aharonov \emph{et al} \cite{Aharonov88,Shikano10,Dresse14} is regarded as a generation of von Neumann measurement, which aims to reveal the odd properties of time-symmetry quantum state \cite{Aharonov64,Vaidman89}. PWM consists of three steps: \emph{i)} initial state preparation for the quantum system and the meter system,\emph{ ii)} weak interaction between them, which can weakly entangle the two systems, but not resolve the shape of the pointer from the previous one, and \emph{iii)} postselection on a final state, which is nearly or absolutely orthogonal to the initial state \cite{orthogonal}. The key consequence of PWM is a large change for the pointer.  This scenario can be used to amplify some tiny interactions \cite{Hosten08,Dixon09}, to directly measure the wave functions of a quantum system \cite{Bamber11}, and to explain some counter-intuitive quantum paradoxes \cite{Aharonov05}.

Pure state is usually used as the measurement pointer in PWM process. Although mixed state is also considered \cite{Johansen04}, the effect induced by postselection on the classical probability distribution (CPD) for each component of the mixed state has not been involved in, because an exact solution to arbitrary pointer is difficultly obtained \cite{Pang14}. In this Letter, for the first time, we presents a scheme to study the effect of postselection on mixed state, which is related to the weak XPM interaction between a single-photon in a superposition state and a thermal light. Here postselection has a classical meaning that CPD (not probability amplitude) of the mixed state can be changed and an amplification effect can be created, which is different from that induced by destructive interference \cite{Aharonov88}. Meanwhile it is also revealed that weak amplification and weak value are not one thing.  After postselection, the classical thermal light become one nonclassical mixed field. This scheme can be realized via the existed techniques.

\emph{Scheme and analysis.} - Fig. 1 is the schematic for our proposal as used in Ref. \cite{Steinberg11,Shikano14}. The quantum system can interact with the meter system through a Kerr medium. The evolution operator of the whole system can be described as exp$(i\phi_{0}\hat{n}_{b}\hat{n}_{c})$, where $\phi_{0}\ll 1$ is the cross-phase shift per photon and $\hat{n}_{b}$ ($\hat{n}_{c}$) is the number operator for mode $b$ ($c$).

 When one single-photon enters a 50:50 beam splitter, it can be prepared initially in an one-photon equal superposition state $|\varphi\rangle=\frac{1}{\sqrt{2}}(|1\rangle_{a}|0\rangle_{b}+|0\rangle_{a}|1\rangle_{b} )$ of the arms $a$ and $b$. If the probe light is prepared in a coherent state $|\alpha \rangle_{c}$ \cite{Steinberg11,Shikano14}, the state of the whole system after interaction is an entangled state $\frac{1}{\sqrt{2}}(|1\rangle_{a}|0\rangle_{b}|\alpha \rangle_{c}+|0\rangle_{a}|1\rangle_{b}|e^{i\phi_{0}}\alpha\rangle_{c})$. The new coherent state $|e^{in\phi_{0}}\alpha\rangle_{c}$ can not be distinguished from $|\alpha \rangle_{c}$, for the phase shift $\phi_{0}$ is  comparatively smaller than the fluctuations $\sigma$ of the coherent light. For observing the weak XPM interaction, postselection strategy is applied to measure the photon in a final state $|\psi\rangle$, which is nearly orthogonal to the initial one $|\varphi\rangle$, that is $\langle \psi |\varphi \rangle\approx 0$. This induces a destructive interference, and generates a very large phase shift on the coherent light to an observed level as a collection of many photons interacting with it. In a word, generally for PWM, postselection, amplification, entanglement and interference are four basic features.

In this Letter, we make use of a thermal light as the pointer rather than the coherent light, which can be described as
\begin{eqnarray}
\rho_{th}(z)=(1-z) \sum_{n=0} z^{n}|n\rangle_{c} \langle n|_{c},
\end{eqnarray}
where $z=e^{-\hbar\omega/k_{B}T}$, $\hbar$ is the Planck constant, $\omega$ is the frequency of the thermal field, $k_{B}$ is the Boltzmann constant, and $T$ is the temperature.  It is clear that $0\leq z <1$. When $\omega$ decrease or $T$ increases, $z$ will increase. After the weak XPM interaction, the quantum system and the meter system will evolve into a mixed (not entangled) state as following
\begin{eqnarray}
\rho_{st} =  (1-z) \sum_{n=0}z^{n}|\varphi_{n}\rangle \langle \varphi_{n}| \otimes
|n\rangle_{c} \langle n|_{c},
\end{eqnarray}
here $|\varphi_{n}\rangle=\frac{1}{\sqrt{2}}(|1\rangle_{a}|0\rangle_{b}+e^{in\phi_{0}}|0\rangle_{a}|1\rangle_{b})$.
For different number state component $|n\rangle_{c}$ in the thermal stat, the weak cross-Kerr nonlinearity can induce different relative phases $n\phi_{0}$ in the single-photon superposition state. For the single-photon, the interaction produces an average phase shift $\bar{n}\phi_{0}$ between arms $a$ and $b$, where $\bar{n}=\frac{z}{1-z}$ is the average photon numbers in the thermal light. Here entanglement as the basic feature for PWM is lost, which is different from that using a coherent field as the probe.

When the single-photon comes out from the second beam splitter and is triggered in the dark port by D1, a final state $|\psi\rangle=\frac{1}{\sqrt{2}}(|1\rangle_{a}|0\rangle_{b}-|0\rangle_{a}|1\rangle_{b} )$ is selected, that is $\langle \psi |\varphi \rangle= 0$ \cite{supplemental}. The final state of the probe light is
\begin{eqnarray}
\rho_{f} = \frac{1-z}{4P}\sum_{n=0}z^{n}(1- e^{in\phi_{0}})(1- e^{-in\phi_{0}})|n\rangle_{c} \langle n|_{c}.
\end{eqnarray}
Here $P=\frac{1}{4}(2-\frac{1-z}{1-ze^{i\phi_{0}}}-\frac{1-z}{1-ze^{-i\phi_{0}}})$ is the post-selected probability in the dark port. A new mixed state is obtained through postselection.

\begin{figure}[b]
\includegraphics[scale=0.14]{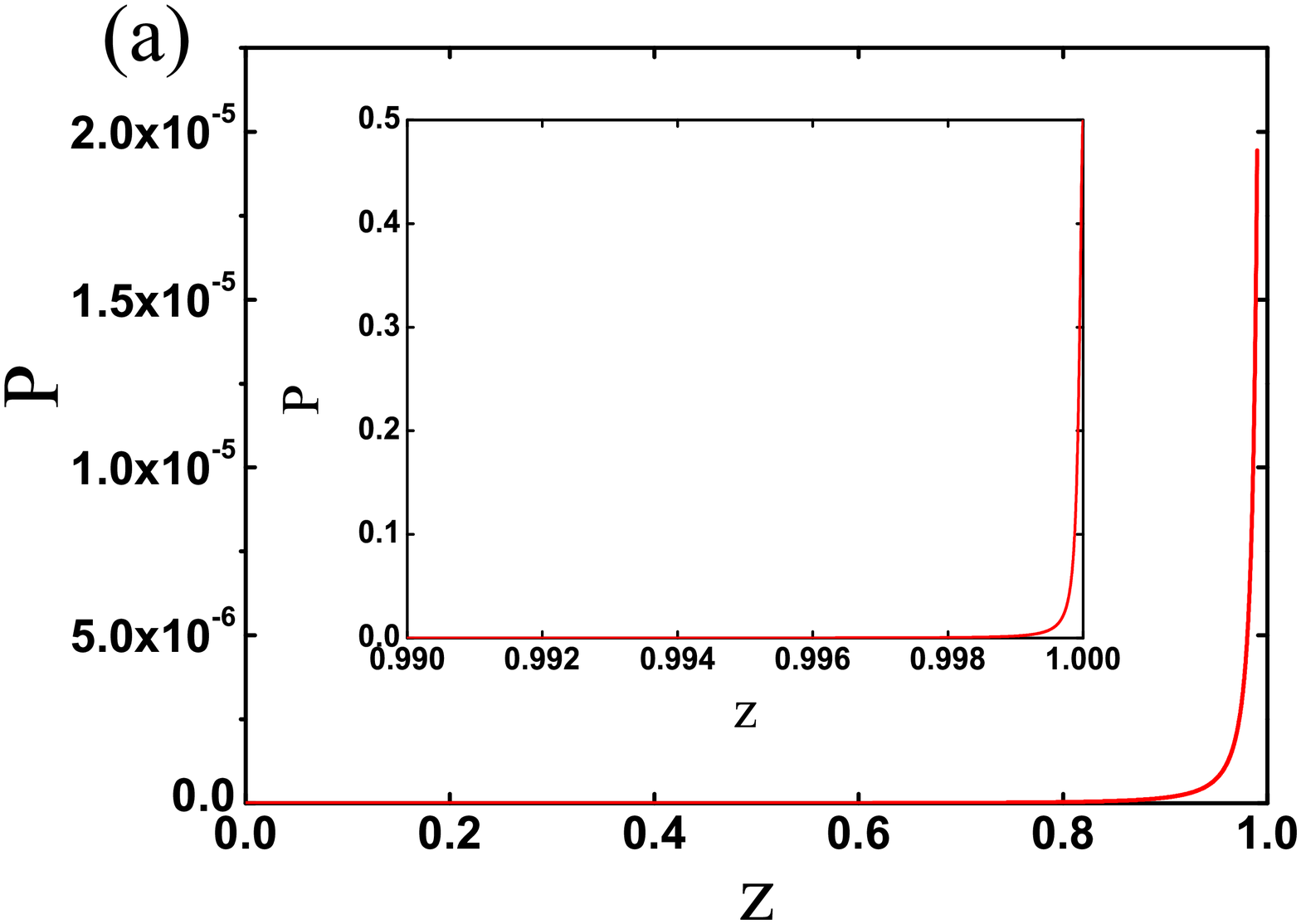}.
\includegraphics[scale=0.14]{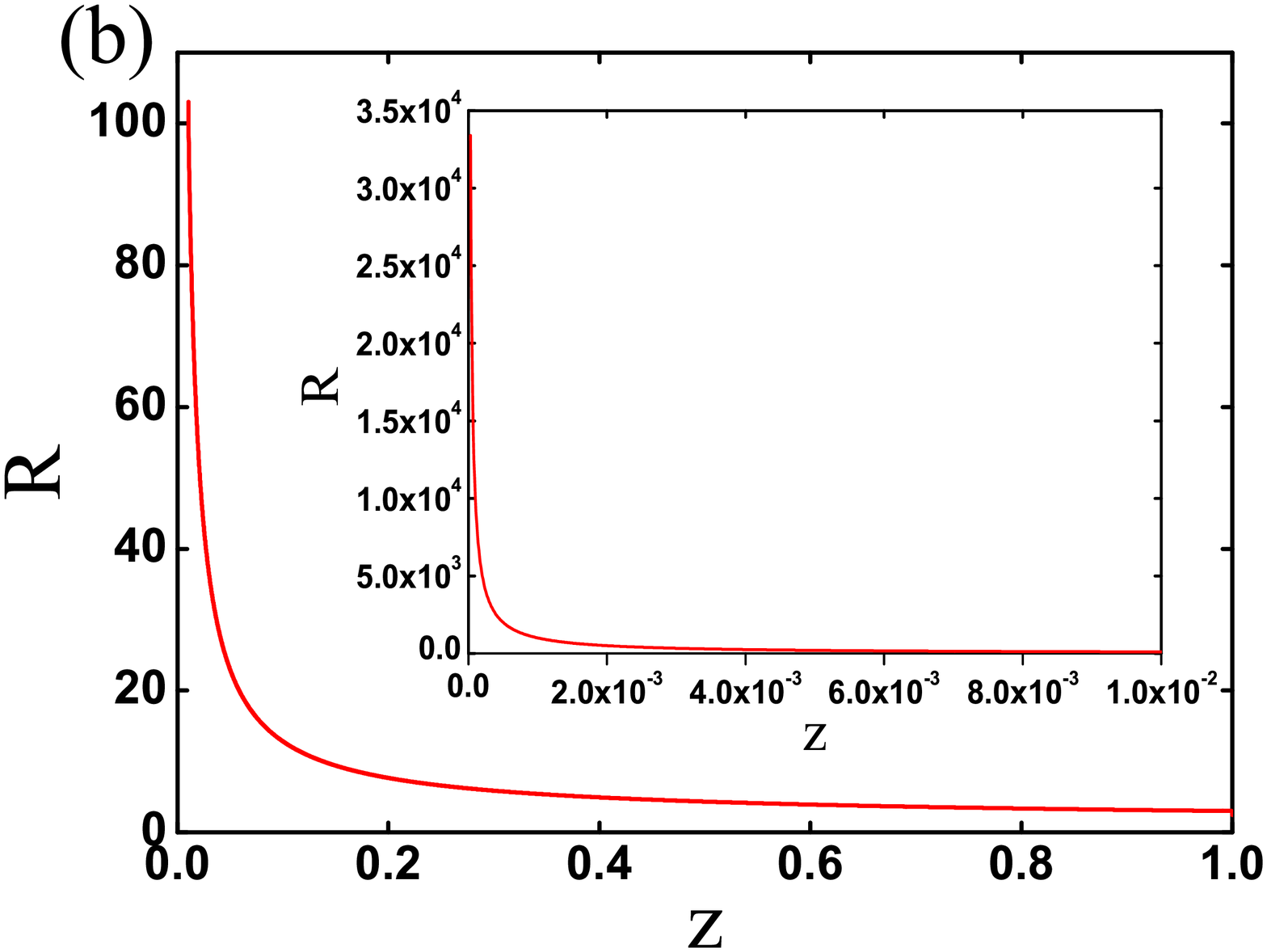}.
\caption{Detection probability $P$ and amplification ratio $R$ for average photon number in the new mixed state and the thermal state are plotted in (a) and (b) as a function of $z$. }
\end{figure}

First of all, we discuss the meaning of the results (3). When $n_{k}\phi_{0}=2k\pi$, and $k=0,1,2\cdot\cdot\cdot$, the success detection probability for the corresponding number states $|n_{k}=\frac{2k\pi}{\phi_{0}}\rangle$ is zero. This means the probe light can not occur in number state $|n_{k}\rangle$ . In fact the disappeared number state can be arbitrarily chosen using a phase shifter \cite{supplemental}. Thus in this Letter, postselection can change the CPD for each number state component $|n\rangle_{c}$ in the mixed state. Here the destructive interference between the postselected meter states does not exist, but we also obtain an amplification effect via measuring the average photon number and the change of the occupation probability.

\begin{figure}[b]
\includegraphics[scale=0.14]{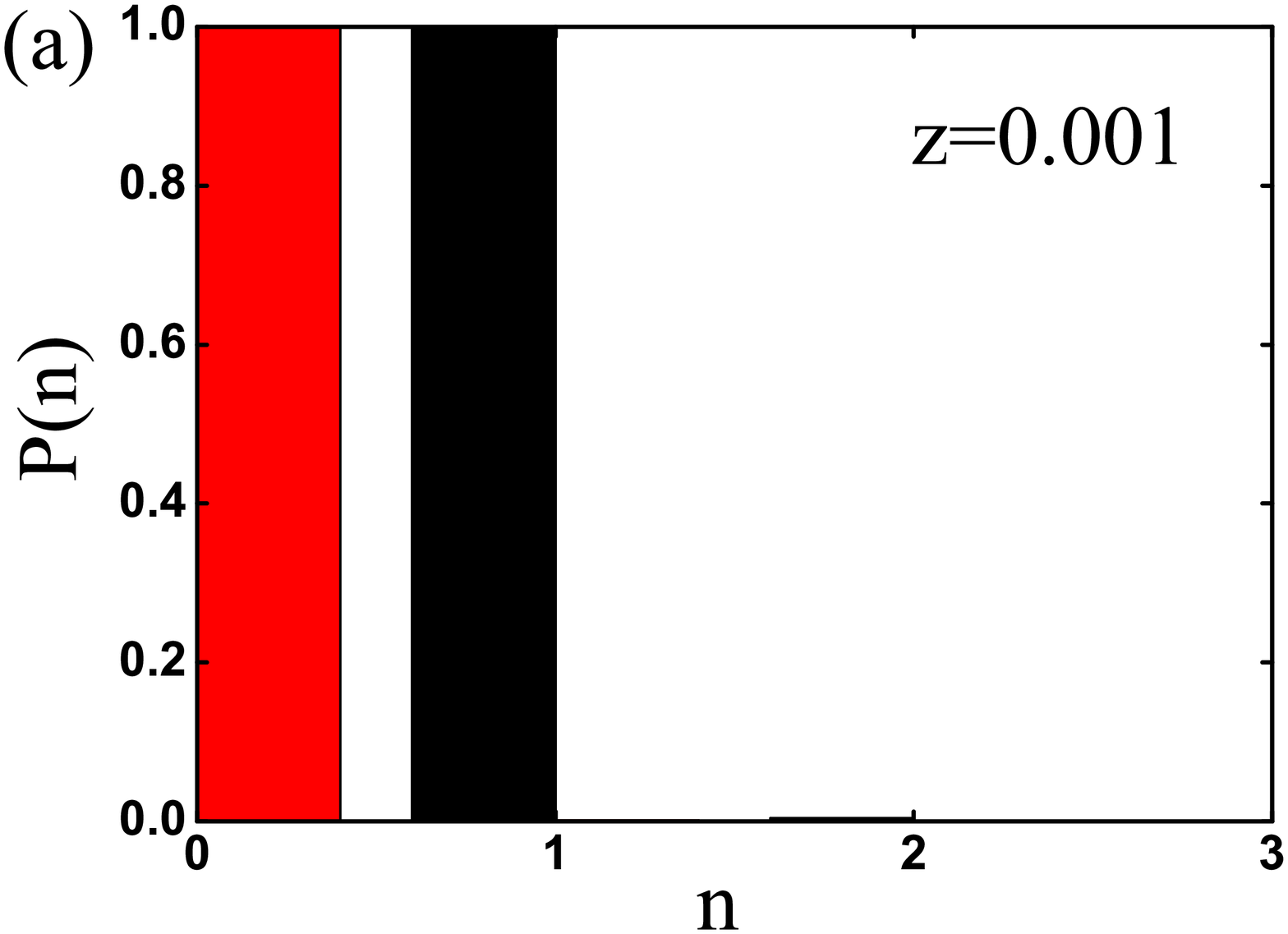}.
\includegraphics[scale=0.14]{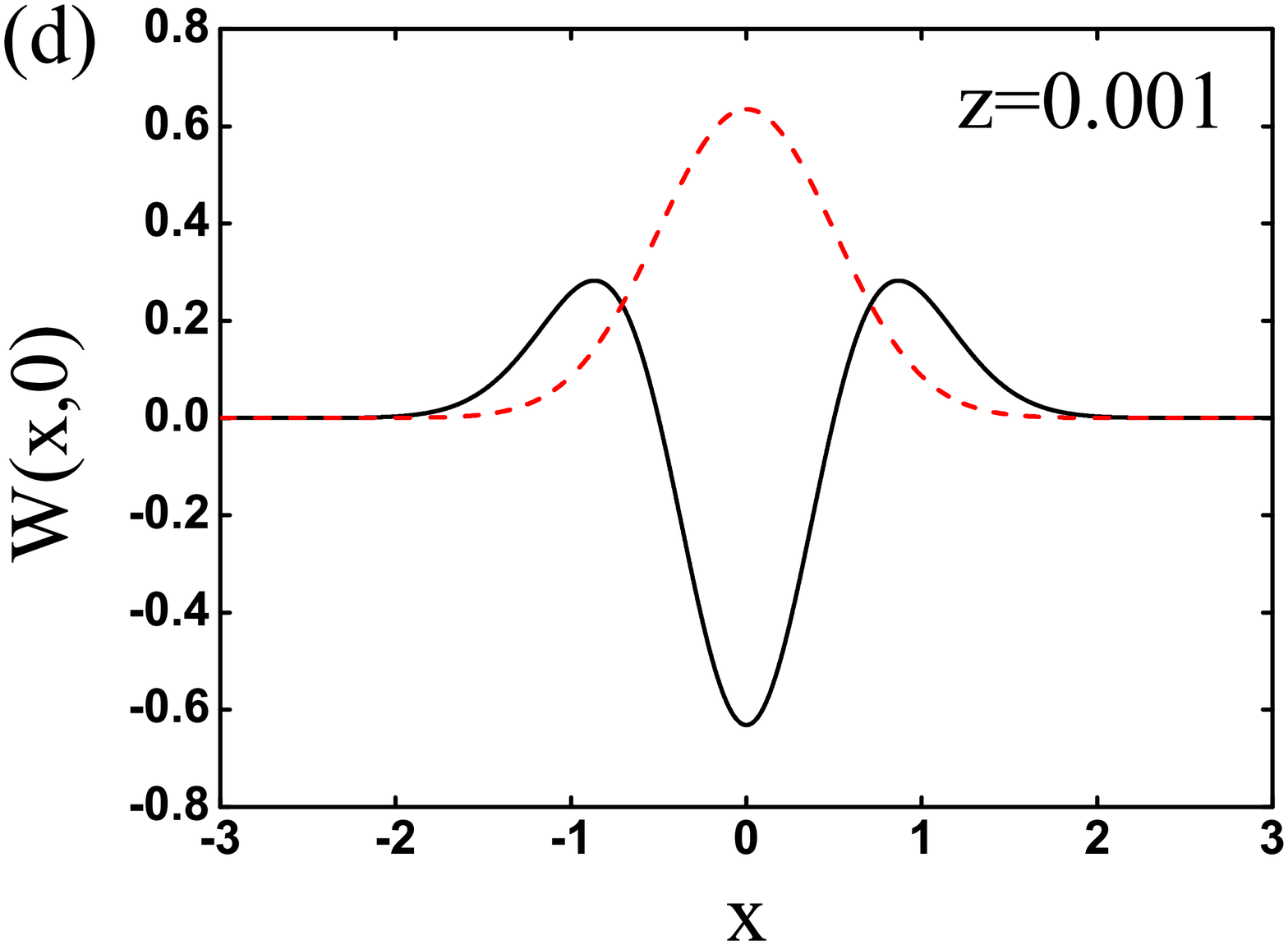}.
\includegraphics[scale=0.14]{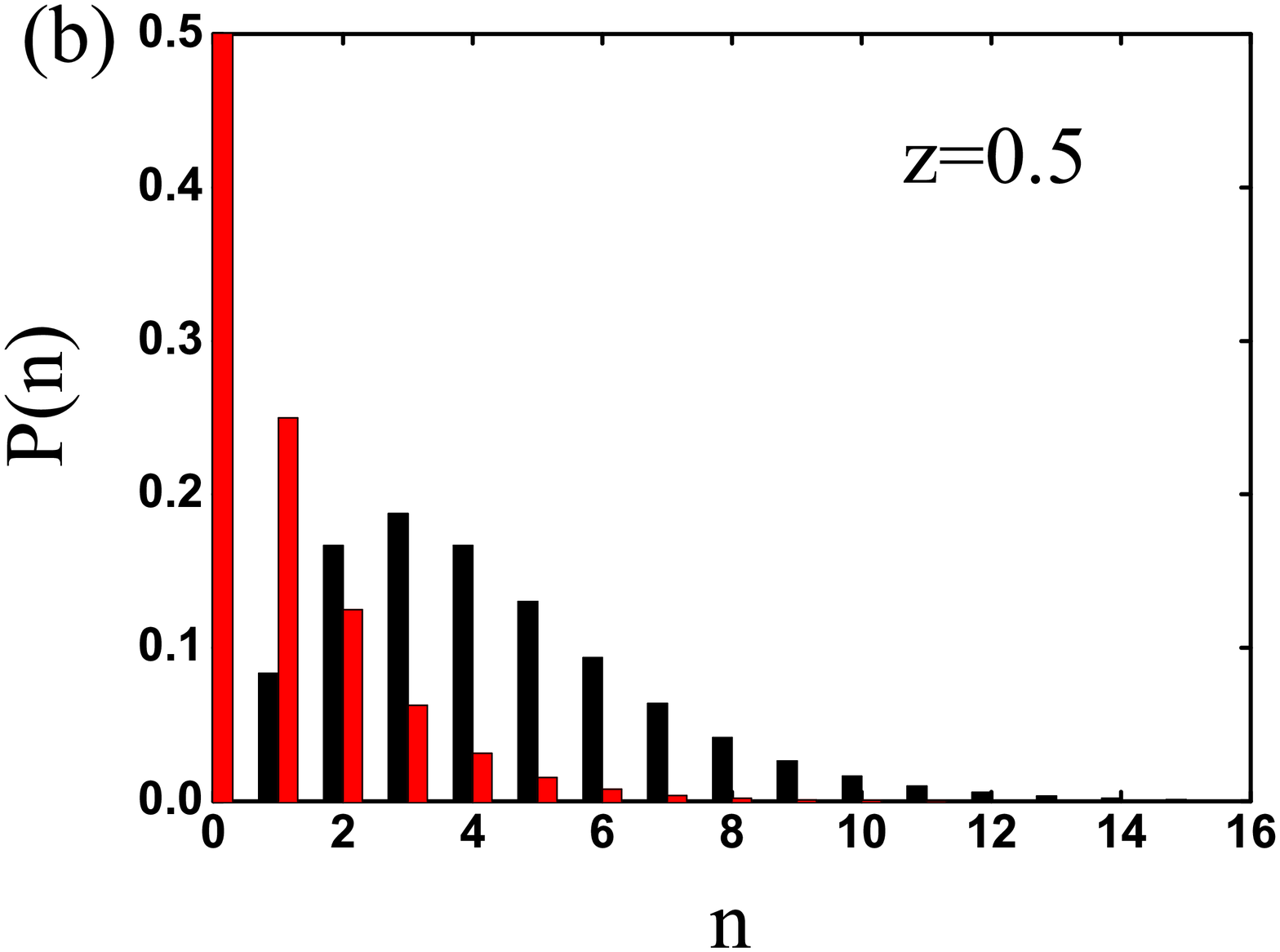}.
\includegraphics[scale=0.14]{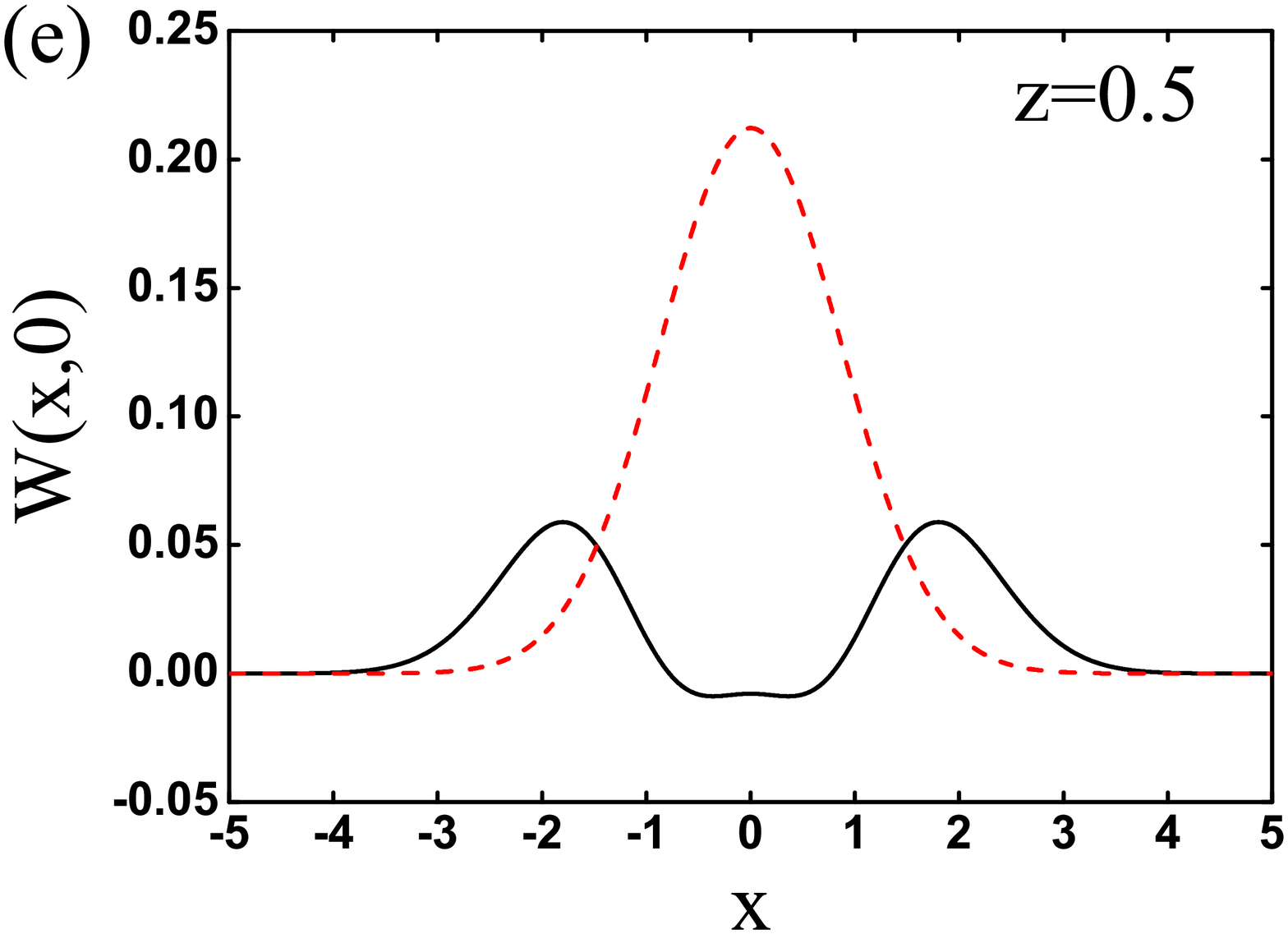}.
\includegraphics[scale=0.14]{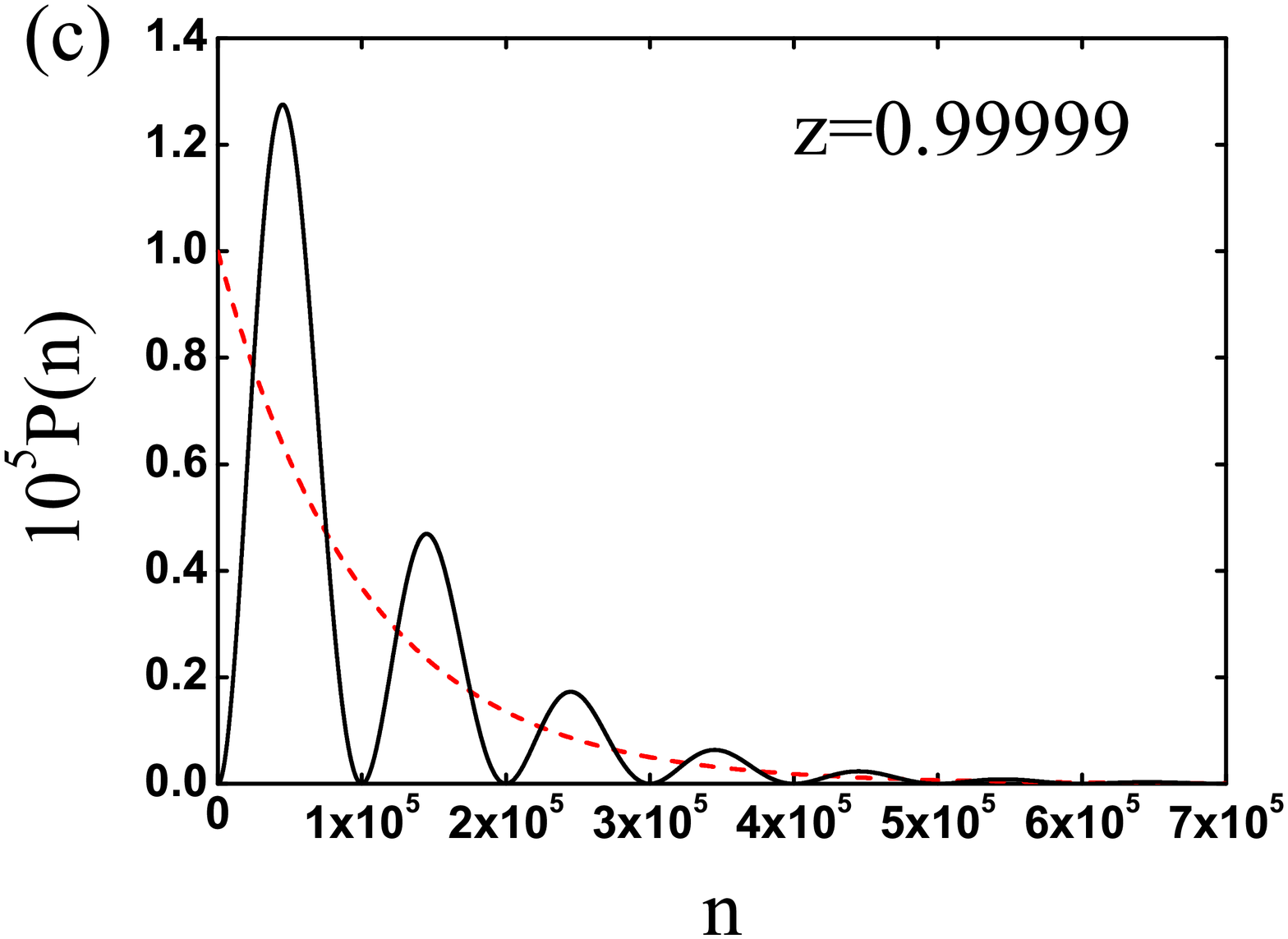}.
\includegraphics[scale=0.14]{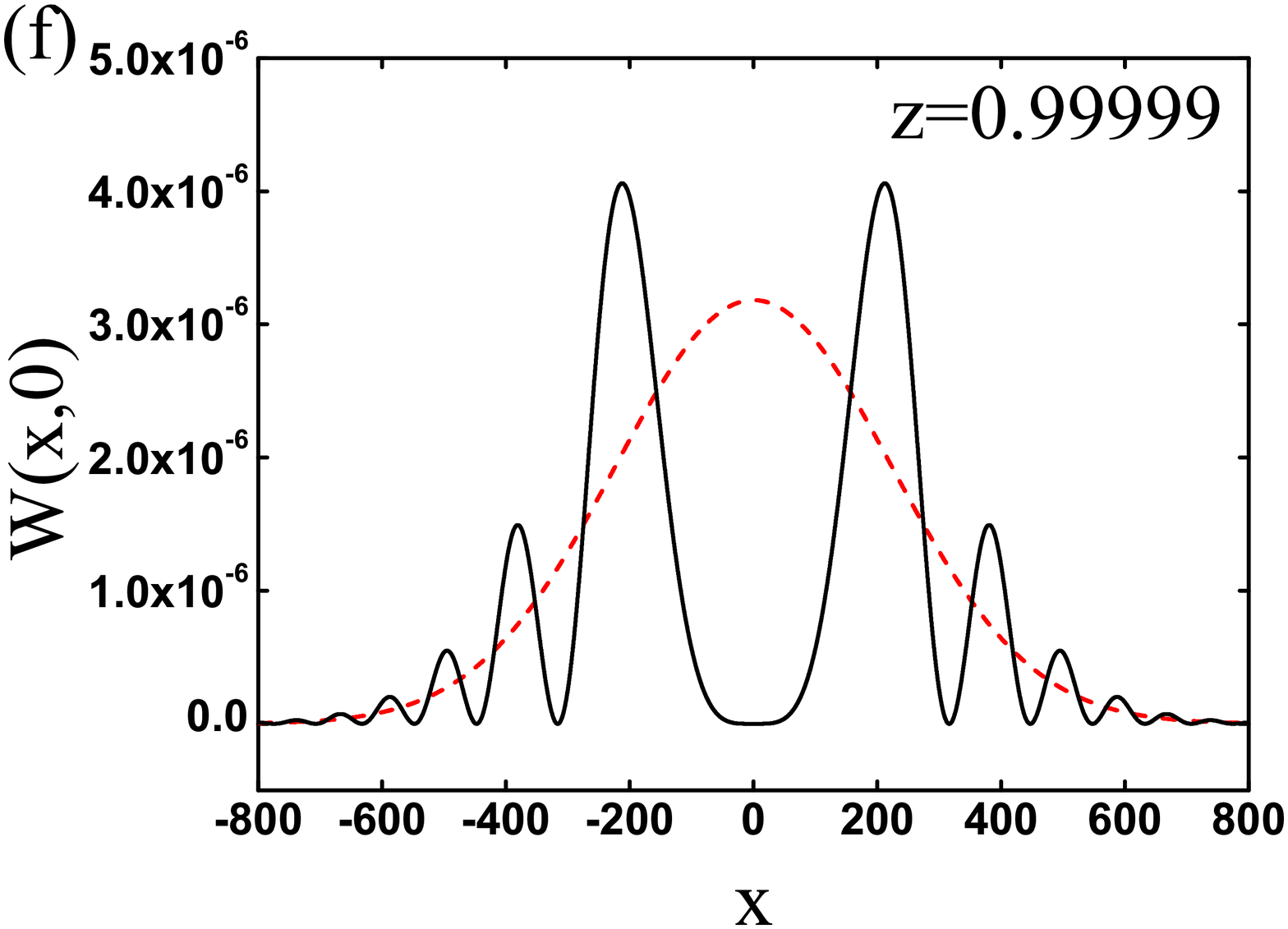}.
\caption{(a),(b) and (c) are the photon number distribution for the generated mixed state (black) and the initial thermal state (red) for $z=0.001$,$0.5$, $0.99999$; (d), (e) and (f) are the corresponding Wigner functions $W(x,0)$.}
\end{figure}

The average photon number of the postselected state is
\begin{eqnarray}
\bar{n}_{f}=\frac{1-z}{4P}\left[\frac{2z}{(1-z)^{2}}-\frac{ze^{i\phi_{0}}}{(1-ze^{i\phi_{0}})^{2}}
-\frac{ze^{-i\phi_{0}}}{(1-ze^{-i\phi_{0}})^{2}}\right]. \notag \\
~~~~~
\end{eqnarray}
We can define a ratio $R=\bar{n}_{f}/\bar{n}$. In Fig. 2 the success probability $P$ and the average number ratio $R$ are plotted. We assume $\phi_{0}=2\pi\times 10^{-5}$ used in \cite{Steinberg11,Shikano14}. It can be seen that when $z$ increases, the success probability also increases, and the amplification ratio $R$ decreases. The smaller the success probability, the larger the amplification ratio.  The amplification effect is very clear. When $z=0.001$, $P=9.9\times 10^{-13}$, $\bar{n}_{f}=1.004$, $R=1003$. When $z=0.5$, $P=2.96\times 10^{-9}$, $\bar{n}_{f}=4.33$, $R=4.33$. When $z=0.99999$, $P=0.4876$, $\bar{n}_{f}=1.05\times 10^{5}$, $R=1.05$.

Next we continue to discuss the properties of the new mixed state (3). In Fig. 3 (a), (b) and (c), the photon number distributions $P(n)$ for $z=0.001, 0.5, 0.99999$ are present. When $z$ is much smaller than $0.1$, the generated state is a single-photon state. It is obvious that the occupation probability distribution of the number state components of the new mixed state is greatly modulated, which demonstrates a prominent amplification effect.

For a deeper understanding of the new state, we further give its Wigner functions (WF) $W(x,p)$, where $x$ and $p$ are two quadratures of the light. Its expression is as following
\begin{eqnarray}
W(x,p)&=&\frac{1-z}{2\pi P}[\frac{2}{1+z}\exp(-\frac{2(1-z)(x^{2}+p^{2})}{1+z}) \notag \\
&&-\frac{1}{1+ze^{i\phi_{0}}}\exp(-\frac{2(1-ze^{i\phi_{0}})(x^{2}+p^{2})}{1+ze^{i\phi_{0}}}) \notag \\
&&-\frac{1}{1+ze^{-i\phi_{0}}}\exp(-\frac{2(1-ze^{-i\phi_{0}})(x^{2}+p^{2})}{1+ze^{-i\phi_{0}}})] \notag \\
~~~
\end{eqnarray}
In Fig. 3 (d), (e) and (f), $W(x,0)$ are plotted for $z=0.001, 0.5, 0.99999$.  When $x=p=0$, we can have
\begin{eqnarray}
W(0,0)=-\frac{z(1-z)^{2}(1-\cos \phi_{0})}{\pi P(1+z)(1+z^{2}+2z\cos \phi_{0})}<0
\end{eqnarray}
so this new mixed state is nonclassical.

When a phase shifter $\theta$ is added in the arm $a$ \cite{supplemental}, another new insight can be obtained. When $\theta$ is very small, this situation corresponds to the traditional weak-value amplification region \cite{Aharonov88}. Amplification ratio $R$ for average photon number in the new mixed state as a function of $\theta$ when $z=0.5$ and $z=0.99999$ are plotted in Fig. 4 (a) and (b). We can see, when $z$ is small, maximum weak amplification occurs in the weak-value amplification region. However when $z$ approaches unitary, the success detection probability approaches 0.5, and maximum amplification is away from the weak-value amplification region. This means weak amplification and weak value are not one thing.

\begin{figure}[b]
\includegraphics[scale=0.14]{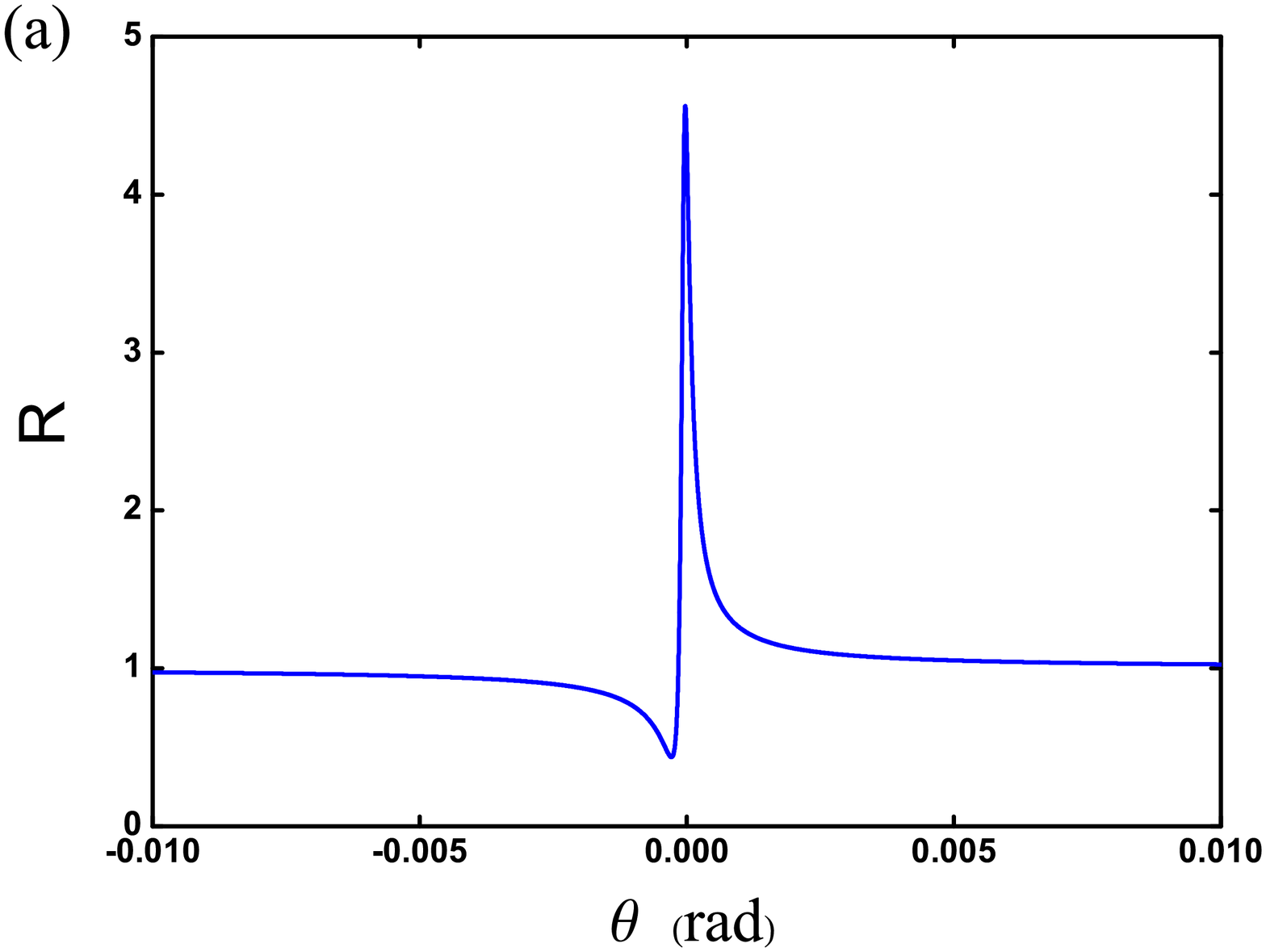}.
\includegraphics[scale=0.14]{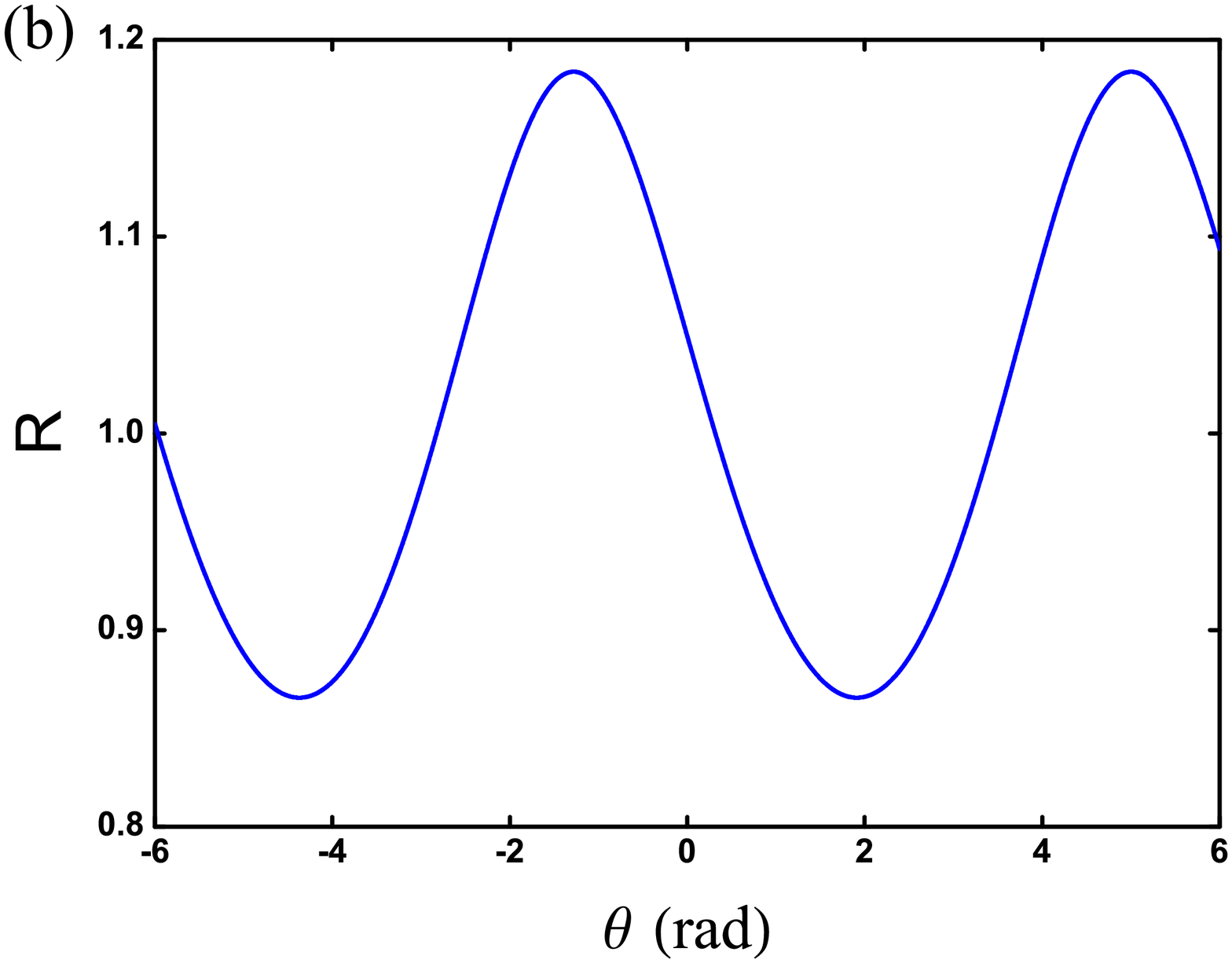}.
\caption{Amplification ratio $R$  as a function of $\theta$ when $z=0.5$ (a) and $z=0.99999$ (b). }
\end{figure}

This Letter presents a classical example for PWM. Although the system is quantum, the amplification can be explained at a classical level. This result is very subtle. Ferrie \emph{et al} showed an evidence \cite{Combes14} that weak values are not inherently quantum, but rather a purely statistical feature of pre- and post-selection with disturbance. This claim arouses much controversies \cite{Vaidman14,Cohen14,Rohrlich14,Sokolovski14,Hofmann14,Brodutch14,Romito15}. Our scheme is very different from that one and satisfies the conditions for PWM. There is no extra disturbance, and the amplification is produced due to only pre- and post-selection. 

Of course this explanation is not the whole story. Thermal state can be diagonal not only in number state basis but also in coherent state basis.  Recently Wang \emph{et al } presents a unified view for PWM \cite{Wang15} bases on the fact that any density matrix could be expanded diagonally in terms of coherent states \cite{Glauber,Sudarshan}. The thermal state can be diagonalized as
\begin{eqnarray}
\rho_{th}(\bar{n})=\frac{1}{\pi \bar{n}}\int d^{2} \alpha e^{-\frac{|\alpha|^{2}}{\bar{n}}}|\alpha\rangle \langle\alpha|.
\end{eqnarray}
This representation is equal to equation (1) but has different physical meaning. Here the thermal light is a mixture of coherent light (classical light). So (2) and (3) can be reexpressed as
\begin{eqnarray}
\rho_{st}&=&\frac{1}{2\pi \bar{n}}\int d^{2}\alpha e^{-\frac{|\alpha|^{2}}{\bar{n}}}(|1\rangle_{a}|0\rangle_{b}|\alpha \rangle_{c}+|0\rangle_{a}|1\rangle_{b}|e^{i\phi_{0}}\alpha\rangle_{c}) \notag \\
&&(\langle 1| _{a} \langle 0| _{b}\langle \alpha|_{c}+\langle 0|_{a}  \langle 1|_{b}\langle e^{i\phi_{0}} \alpha|_{c}),
\end{eqnarray}
and
\begin{eqnarray}
\rho_{f} &=& \frac{1}{4\pi \bar{n} P}\int d^{2}\alpha e^{-\frac{|\alpha|^{2}}{\bar{n}}}(|\alpha\rangle_{c}-|e^{i\phi_{0}}\alpha\rangle_{c}) \notag \\
&& (\langle\alpha |_{c}-\langle e^{i\phi_{0}}\alpha|_{c}),
\end{eqnarray}
so the (classical) interference is in fact averaged out \cite{supplemental}!

\emph{Possibility of experimental realization.} - PossibaWeak XPM in this scheme at the single-photon level has been realized by the established methods in \cite{Gaeta13} and \cite{Steinberg15}. Especially in \cite{Steinberg15,theory} weak XPM has been directly observed at the single-photon level with assistance of electromagnetically-induced transparency (EIT) \cite{EIT}. XPM based on EIT has been studied in \cite{GiantKerr,Zhu,Chen1,Chen2, Lukin}. A pure
cross-Kerr interaction can be generated in $N$-type EIT medium \cite{effevtive1,effevtive2}, and  thermal light EIT has been investigated in \cite{Kim}.

Weak thermal pulses in the spectral region around 800 nm can be generated  by inserting a rotating ground glass disk in the path of the laser beam \cite{Arecchi65}, which was used in \cite{Bellini07} to adding  and  subtracting single photons to and from a thermal field. The quadrature distributions for the new mixed state can be obtained experimentally, and the corresponding photon number probability distributions and  WF can be reconstructed from the experimental data by means of an iterative maximum likelihood algorithm \cite{Sacchi99,Lvovsky04,Rehacek06,Raymer09}. The negativity of the WF is a good indication of the nonclassical character of the new mixed state. A recycled method can be used to enhance the detection efficiency at the dark port \cite{recycle1,recycle2}. This method can be also applied in \cite{Steinberg11,Shikano14} although they did not mention it. When the photon comes from the bright port the corresponding probe light also escapes from XPM, so this photon can again enter into the setup. If $z$ approaches to unity, the nonclassical light can be generated with nearly half the chance which may be used in quantum information process., The single-photon entangled state can be also prepared with the method in Ref \cite{Silberberg} besides exploiting the single-photon source.

In this Letter, we deal with the single-photon and the thermal light in the single-mode approximation. In realistic treatment, continuous-mode photonic pulses should be considered, which makes two individual photons via XPM inseparable \cite{BingHe11}. A realistic picture of quantum phase gate performance was clarified in \cite{BingHe12}. Whether a single-photon and a thermal light can be entangled in the continuous-mode treatment is still not clear. If it is, the real amplification effect will show a small deviation from that in this scheme.  It will be discussed in next paper.

\emph{Conclusions.} - an amplification scheme for PWM with classical explanation is presented. Postselection can induce a large change  on the probability distribution of the mixed state. In this view, the destructive interference between the postselected final states of the meter system does not exist. We explain the result from a coherent-state basis view that the interference is in fact averaged out. In this Letter we find that the effect of postselection is at a classical level.  However the initial superposition state of the quantum system is essential for this scheme. How much classical PWM can be is an important and interesting issue. Meanwhile we point out that weak amplification and weak value are not one thing.

Here thermal light cross-Kerr effect is predicted for the first time.  A new mixed state can be generated with nonclassicality. The new effect has potential applications in quantum information process. This scheme can be realized based on the present techniques. Of course this new effect can be also applied to other quantum systems with XPM interaction, such as optomechanical system.

\emph{Note added}. - During the completion of this work, we became aware of a related work by Mart\'{\i}nez-Rinc\'{o}n, Liu, Viza and Howell \cite{Howell15}. In their and our papers, postselection plays the classical role. However they goes further and re-selectes new final states of the quantum system. Through the new postselection, we can see clearly that weak amplification and weak value are not one thing. The relationships and generalization of the results in the two papers are clarified in \cite{postselection}.

We thank Bing He for helpful comments on weak XPM. This research is supported by National Natural Science
Foundation of China (Grant No.11174109) and (Grant No. 11404242).

~~\\
~~\\

Supplementary materials:

~~\\
(1) Phase shifter $\theta$
~~\\

 We should notice that, if we add a phase shifter $\theta$ in the arm $a$, the initial state of the photon will be in $|\psi(\theta)\rangle=\frac{1}{\sqrt{2}}(e^{i\theta}|1\rangle_{a}|0\rangle_{b}+|0\rangle_{a}|1\rangle_{b} )$. When the final state of the single photon
$|\psi\rangle=\frac{1}{\sqrt{2}}(|1\rangle_{a}|0\rangle_{b}-|0\rangle_{a}|1\rangle_{b} )$ is selected, the final state of the probe light is
\begin{eqnarray}
\rho_{f}(\theta) &=& \frac{1-z}{4P'}\sum_{n=0}z^{n}(1- e^{i(\theta+n\phi_{0})})  \notag \\
&&(1- e^{-i(\theta+n\phi_{0})})|n\rangle_{c} \langle n|_{c},
\end{eqnarray}
where $P'=\frac{1}{4}(2-\frac{(1-z)e^{i\theta}}{1-ze^{i\phi_{0}}}-\frac{(1-z)e^{-i\theta}}{1-ze^{-i\phi_{0}}})$. When $\theta+n_{k}\phi_{0}=2k\pi$, $k=0,1,2,\cdot\cdot\cdot$, the distribution probability for number state $|n_{k}\rangle$ is zero. So we can eliminate the number state component $|n\rangle_{c}$ via adjusting the phase shifter $\theta$.

~~\\
(2) The single-photon is initially prepared in $|1\rangle_{b}$
~~\\

For a pure number state $|n\rangle_{c}$ in mode $c$, although it rotates an angle $n\phi_{0}$ after the weak XPM interaction with a single-photon, the state is also the same number state $|n\rangle_{c}$. That is exp$(i\phi_{0}\hat{n}_{b}\hat{n}_{c})|1\rangle_{b}|n\rangle_{c}=e^{in\phi_{0}}|1\rangle_{b}|n\rangle_{c}$. So if the single-photon is initially prepared in $|1\rangle_{b}$ (single mode approximation), the thermal light can not change. In the realistic situation interactions
can generically create continuous-mode entanglement \cite{BingHe11}. However in our Letter the classical explanation of postselection relies on the single-mode approximation.

~~\\
(3) interference average
~~\\

\begin{eqnarray}
\rho_{f} &=& \frac{1}{4\pi \bar{n} P}\int d^{2}\alpha e^{-\frac{|\alpha|^{2}}{\bar{n}}}(|\alpha\rangle_{c}-|e^{i\phi_{0}}\alpha\rangle_{c})
(\langle\alpha |_{c}-\langle e^{i\phi_{0}}\alpha|_{c}) \notag \\
&=& \frac{1}{4\pi \bar{n} P}\int d^{2}\alpha e^{-\frac{|\alpha|^{2}}{\bar{n}}}(e^{-\frac{|\alpha|^{2}}{2}}\sum_{n=0} \frac{\alpha^{n}}{\sqrt{n!}}|n\rangle_{c}-e^{-\frac{|\alpha|^{2}}{2}}\sum_{n=0} \frac{e^{in\phi_{0}}\alpha^{n}}{\sqrt{n!}}|n\rangle_{c}) \notag \\
&&(e^{-\frac{|\alpha|^{2}}{2}}\sum_{m=0} \frac{\alpha^{*m}}{\sqrt{m!}}\langle m|_{c}-e^{-\frac{|\alpha|^{2}}{2}}\sum_{m=0} \frac{e^{-im\phi_{0}}\alpha ^{*m}}{\sqrt{m!}}\langle m|_{c})  \notag \\
&=& \sum_{n=0}\sum_{m=0} ( \frac{1}{4\pi \bar{n} P}\int d^{2}\alpha \frac{\alpha^{n}\alpha^{*m} }{\sqrt{n!} \sqrt{m!}}e^{-\frac{\bar{n}+1}{\bar{n}}|\alpha|^{2}} )(1- e^{in\phi_{0}})(1- e^{-im\phi_{0}})|n\rangle_{c} \langle m|_{c} \notag \\
&=&\sum_{n=0}\frac{1-z}{4P}z^{n}(1- e^{in\phi_{0}})(1- e^{-in\phi_{0}})|n\rangle_{c} \langle n|_{c},
\end{eqnarray}
so for thermal state, the interference is averaged out at last. This result can be generalized to any diagonal mixed state in the number state basis.


\begin{thebibliography}{99}
\bibitem{orthogonal} Generally speaking, when the final state is nearly orthogonal to the initial state we can have an amplification effect, and if it is absolutely orthogonal we can have a more different state from the initial one and no amplification effect. However these two situations are not hold in the Letter.
\end{thebibliography}
\end{document}